\documentclass[aps,prl,twocolumn,superscriptaddress]{revtex4-2}

\usepackage{amssymb}
\usepackage{amsmath}
\usepackage{graphicx}
\usepackage{array}
\usepackage{multirow}
\usepackage{setspace}
\usepackage[colorlinks,linkcolor=blue,anchorcolor=blue,citecolor=red]{hyperref}

\begin{document}

\title{Mechanism of magnetic phase transition in correlated magnetic metal: insight into itinerant ferromagnet Fe$_{3-\delta}$GeTe$_2$}
\author{Yuanji Xu}
\affiliation{Institute for Applied Physics, University of Science and Technology Beijing, Beijing 100083, China}
\author{Yue-Chao Wang}
\affiliation{Laboratory of Computational Physics, Institute of Applied Physics and Computational Mathematics, Beijing 100088, China}
\author{Xintao Jin}
\affiliation{Institute for Applied Physics, University of Science and Technology Beijing, Beijing 100083, China}
\author{Haifeng Liu}
\affiliation{Laboratory of Computational Physics, Institute of Applied Physics and Computational Mathematics, Beijing 100088, China}
\author{Yu Liu}
\affiliation{Laboratory of Computational Physics, Institute of Applied Physics and Computational Mathematics, Beijing 100088, China}
\author{Haifeng Song}
\email{song$\_$haifeng@iapcm.ac.cn}
\affiliation{Laboratory of Computational Physics, Institute of Applied Physics and Computational Mathematics, Beijing 100088, China}
\author{Fuyang Tian}
\email{fuyang@ustb.edu.cn}
\affiliation{Institute for Applied Physics, University of Science and Technology Beijing, Beijing 100083, China}
\date{\today}

\begin{abstract}
Developing a comprehensive magnetic theory for correlated itinerant magnets poses challenges due to the difficulty in reconciling both local moments and itinerant electrons. In this work, we investigate the microscopic process of magnetic phase transition in ferromagnetic metal Fe$_{3-\delta}$GeTe$_2$. We find that Hund's coupling is crucial for establishing ferromagnetic order. During the ferromagnetic transition, we observe the formation of quasiparticle flat bands and an opposing tendency in spectral weight transfer, primarily between the lower and upper Hubbard bands, across the two spin channels. Moreover, our results indicate that one of the inequivalent Fe sites exhibits Mott physics, while the other Fe site exhibits Hund's physics, attributable to their distinct atomic environments. We suggest that ferromagnetic order reduces spin fluctuations and makes flat bands near the Fermi level more distinct. The hybridization between the distinctly flat bands and other itinerant bands offers a possible way to form heavy fermion behavior in ferromagnets. The complex interactions of competing orders drive correlated magnetic metals to a new frontier for discovering outstanding quantum states.
\end{abstract}

\maketitle

\vspace{10pt}
\noindent \textbf{Introduction} \\
Magnetism plays a crucial role in condensed matter physics, especially in strongly correlated systems. It is closely linked to various emergent phenomena such as heavy fermion behavior \cite{Stewart1984,Yang2008}, unconventional superconductivity \cite{Dagotto1994,Scalapino2012}, Hund metal \cite{Yin2011,Haule2009}, and quantum phase transition \cite{Si2001,Shen2020}. These fascinating behaviors arise due to the dual nature of correlated electrons, which creates challenges in understanding the role of magnetism. One essential scenario is the delocalization of $f$-electrons at low temperatures in heavy fermion systems, achieved through the hybridization of local moments with conductive electrons via RKKY interaction \cite{Shim2007}. Another critical instance is the $d$-electron Hund metals, which appear even more mysterious as the systems do not have explicit two fluids \cite{Georges2024}. Unlike Mott physics, which is characterized by significant on-site Coulomb repulsion $U$ that impedes charge fluctuations and can lead to electron localization, Hund physics is more influenced by Hund's coupling $J$. This coupling encourages electrons to adopt a collective configuration with maximum spin. The odd coexistence of itinerant and localized behaviors in Hund metals, dominated by Hund's coupling, has scored significant victories in iron-based superconductors \cite{Yin2011,Georges2013,Haule2008,Yin22011,Liu2012}. However, from a theoretical perspective, the two commonly used models, one starting from the Heisenberg local moment model and the other from the Stoner itinerant electron model \cite{Heisenberg1928,Stoner1947}, have their limitations in treating these systems. The microscopy mechanism of considering both local moments and itinerant electrons in correlated magnetic metal still needs to be explored.

Recently, van der Walls (vdW) itinerant ferromagnet Fe$_{3}$GeTe$_2$ (FGT) has garnered significant attention. It is a potential candidate for voltage-controlled spintronic devices due to its exceptional exfoliation properties and robust high Curie temperature \cite{Deng2018,Fei2018}. In bulk material, the ferromagnetic (FM) Curie temperature $T_{c}$ is around 220 K, and it can even reach room temperature with an ionic gate in its thin flakes \cite{Deng2018,Deiseroth2006,Chen2013}. Furthermore, experiments have observed significant electron mass enhancement \cite{Chen2013,Verchenko2015,Zhang2018,Zhao2021}, which is also supported by previous theoretical calculations \cite{Zhu2016}. Considering its high ferromagnetic transition temperature and strongly correlated effects, the itinerant ferromagnet FGT presents an ideal platform for investigating the dual nature of magnetism and its interplay with other exotic phenomena \cite{Kim2018}.

Here, we highlight some mysteries that may be related to the issue of magnetism in FGT. First and foremost is the question of how to describe the Fe-$d$ electrons accurately. Although the itinerant Stoner model is commonly accepted \cite{Chen2013,Zhuang2016}, recent experiments suggest that the Heisenberg model may be necessary to describe the ferromagnetism in FGT correctly \cite{Xu2020,Corasaniti2020,Bao2022,Wu2024}. For instance, the angel-resolved photoemission spectroscopy (ARPES) experiments find that the temperature-dependent evolution towards the ferromagnetic transition is strikingly weak \cite{Xu2020,Wu2024}. Secondly, as is commonly discussed in iron-based superconductors \cite{Yin2011,Song2016,Deng2019}, it is unclear whether the irons in FGT exhibit Hund physics or Mott physics. Kim et al. suggest that it is a ``site-differentiated" Hund metal \cite{Kim2022}, while another study suggests that it is close to the orbital selective Mott physics \cite{Bai2022}. Finally, previous studies have reported heavy fermion behavior in FGT with a large Sommerfeild coefficient ($\gamma_0 = 135\,$mJ/mol K$^2$) \cite{Chen2013,Verchenko2015,Zhang2018}. This is striking since $d$-electron heavy fermion systems are rare \cite{Kondo1997,Cheng2013,Takegami2022}. While in FGT, heavy fermion behavior appears to be promoted within ferromagnetic order \cite{Zhang2018}. These controversies encourage us to explore the complex interaction of magnetism with possible Hund, Mott, or Kondo physics in this system.

In this work, we systematically investigate the correlation effects in itinerant ferromagnet Fe$_{3-\delta}$GeTe$_2$ using density functional theory combined with dynamical mean-field theory (DFT+DMFT). Our findings reveal a detailed evolution of electronic structures during magnetic transition in ferromagnets, characterized by spectral weight transfer between the two spin channels. This process is driven by dynamical correlation effects with frequency dependence self-energy, rather than by static spin splitting with band shift in the mean-field Stoner model. Similarly to Hund metals, the quasiparticles form in itinerant ferromagnet as temperature decreases. Interestingly, the Fe atoms in FGT exhibit two distinct behaviors: one follows Mott physics, while the other exhibits Hund physics. This suggests that FGT may be the material to exhibit both Hund and Mott physics, stemming from the multi-site nature of correlated atoms in its crystal structure. Furthermore, our results indicate that Hund’s coupling promotes the formation of ferromagnetic order, which in turn reduces spin fluctuations and enhances the clarity of the flat bands near the Fermi level. These well-defined flat bands may subsequently hybridize with other itinerant bands, leading to heavy fermion behavior at lower temperatures, akin to what is observed in $f$-electronic heavy fermion materials. Our findings not only naturally reconcile the conflicts between heavy fermion behavior and long-range magnetic order, but also open up a new frontier for discovering new quantum states in itinerant magnets.

\begin{figure}
\begin{center}
\includegraphics[width=0.50\textwidth]{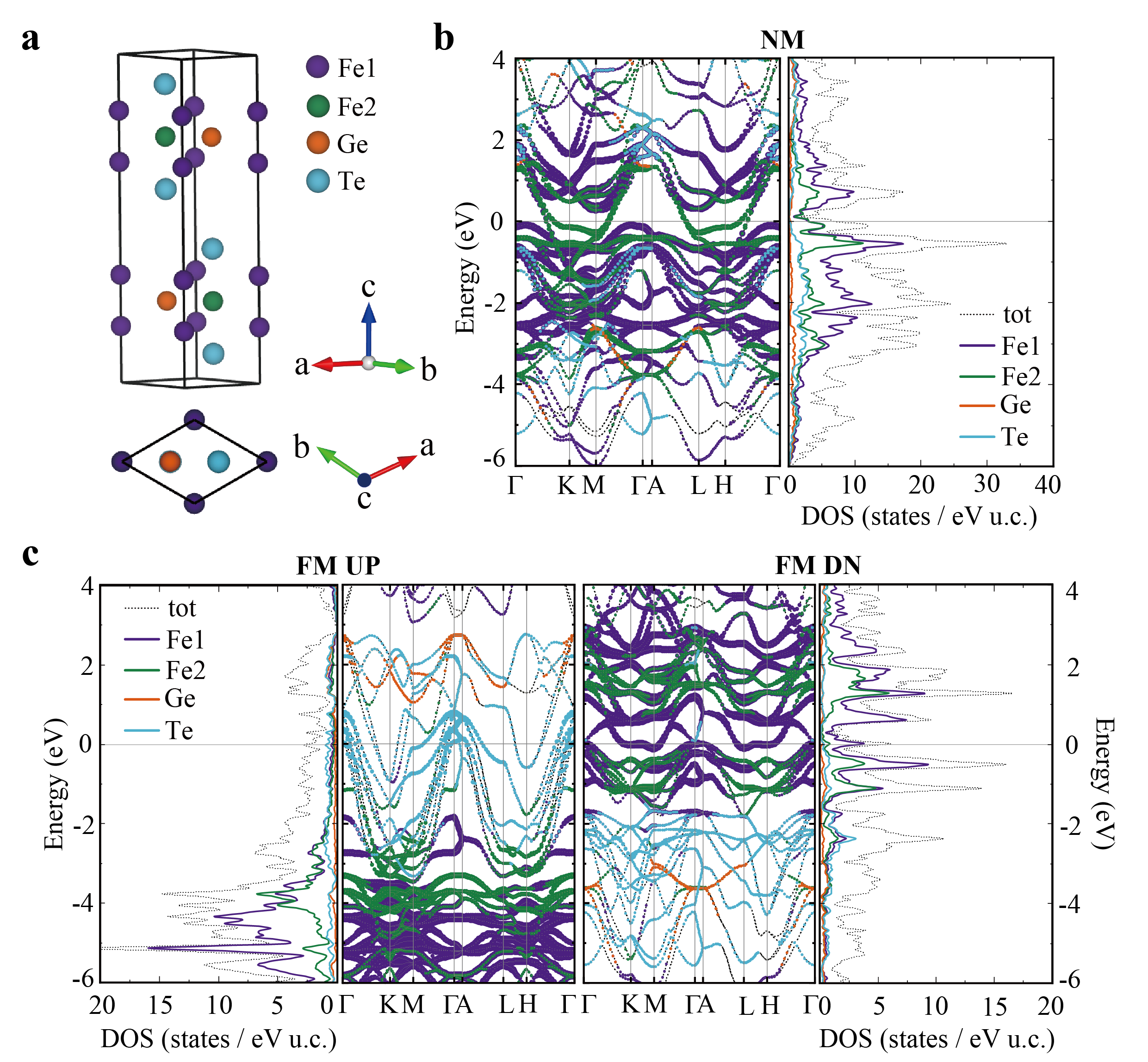}\caption{ \textbf{Crystal structure and DFT+U electronic structures in FGT.} \textbf{a} Illustration of the crystal structure of Fe$_{3}$GeTe$_{2}$. \textbf{b} The electronic structures and partial density of states of non-spin polarized calculations of Fe$_{3}$GeTe$_{2}$ in DFT+U calculations. \textbf{c} The electronic structures and partial density of states of ferromagnetic order in DFT+U calculations.}
\label{fig1}
\end{center}
\end{figure}

\vspace{10pt}
\noindent \textbf{Results} \\
\textbf{DFT+U calculations}\\
Fe$_{3-\delta}$GeTe$_{2}$ is a ternary multi-site ferromagnet, as illustrated in Fig.~\ref{fig1}a, with layered Fe$_{3}$Ge substructures sandwiched by two layers of Te atoms and a vdW gap between adjacent Te layers. FGT crystallizes into a hexagonal structure with the space group of $P6_{3}/mmc$ (No. 194), and Fe atoms occupy two inequivalent Wyckoff sites, labeled Fe1 and Fe2. Notably, the occupational deficiency ($0 \textless \delta \textless 0.3$) occurs only at the Fe2 site \cite{Deiseroth2006,Verchenko2015}. Our DFT results are consistent with available calculations (See Supplementary Note 1) \cite{Verchenko2015,Bai2022}. Considering the strong correlation effects in FGT, we perform DFT+U calculations using the correlation values suggested by previous work \cite{Zhu2016}. 

In non-magnetic (NM) DFT+U calculations, the Fe-$d$ electrons dominate near the Fermi level in Fig.~\ref{fig1}b. However, in ferromagnetic order, the correlation effects strongly localize Fe-$d$ electrons, causing the Fe-bands to be pushed away from the Fermi level. As a result, the Fe-bands in the spin-up channel are almost entirely filled and pushed down to 5 eV below the Fermi level, leaving only itinerant bands with Te character in Fig.~\ref{fig1}c. In contrast, the bands in the spin-down channel move up slightly, leaving a small peak of Fe-$d$ density of states (DOS) at the Fermi level. Besides, the partial DOS of Fe1 and Fe2 have similar peak structures, indicating strong hybridization between them. We note that the DOS at the Fermi level in non-magnetic GGA+U calculations is reduced to 2.9 states/eV f.u., much smaller than 13.6 in GGA calculations \cite{Verchenko2015}. Moreover, the calculated band structures in the ferromagnetic state exhibit a large exchange splitting of over 4 eV compared to the non-magnetic state. However, ARPES measurements indicate that upon heating across the Curie temperature, only minor changes in the shape and position of the band dispersions are observed \cite{Xu2020,Wu2024}. Hence, the much smaller DOS in DFT+U calculations, combined with the weak temperature-dependent evolution of Fe-$d$ bands in ARPES measurements, suggests to reconsider the application of Stoner model to FGT.

\begin{figure*}
\begin{center}
\includegraphics[width=0.98\textwidth]{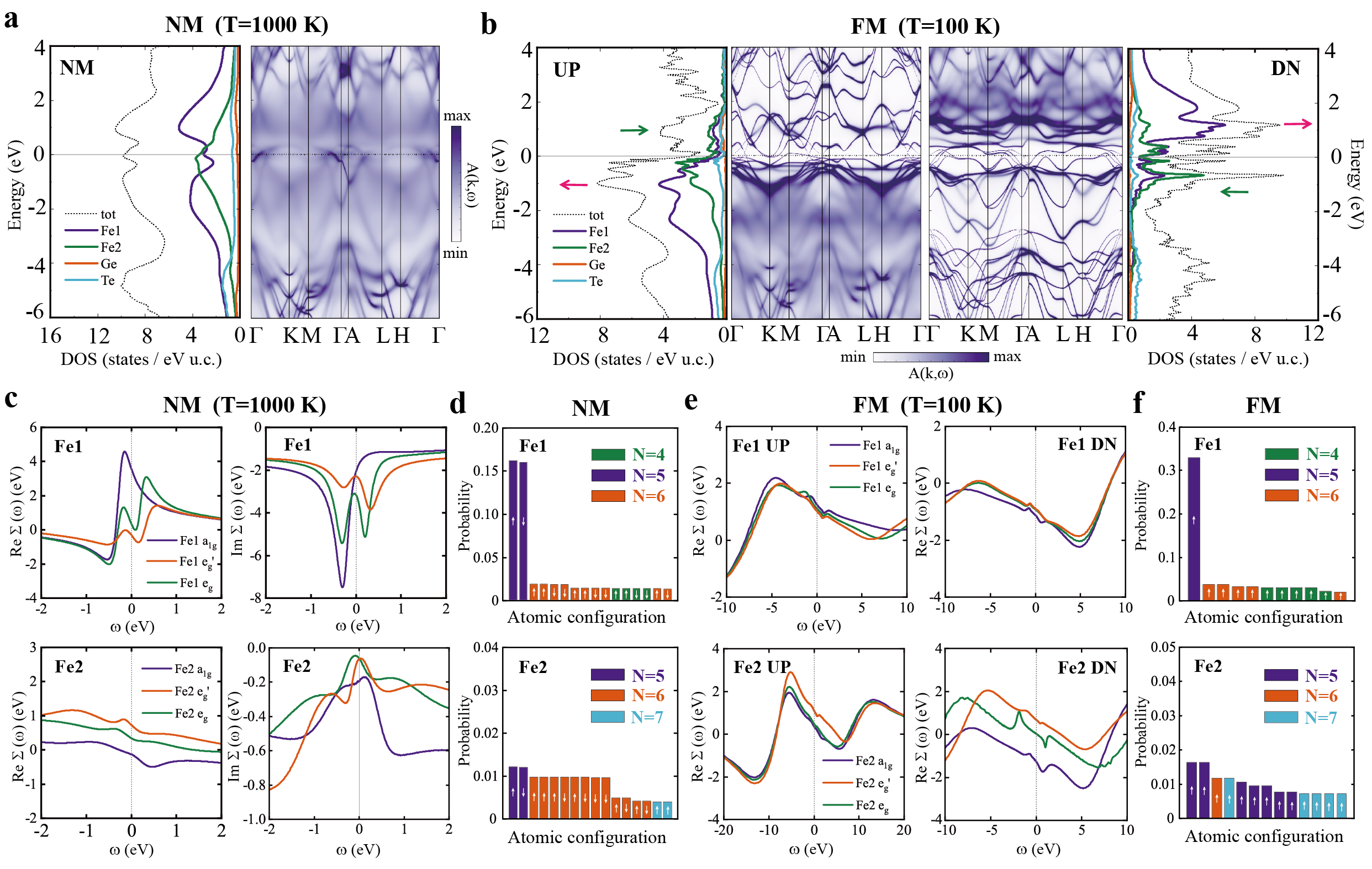}
\caption{\textbf{The results of DFT+DMFT calculations of FGT at low and high temperatures.} \textbf{a,b} The calculated density of states and spectral functions in high-temperature non-magnetic order and low-temperature ferromagnetic order.  In ferromagnetic order, the pink arrow in DOS shows the region of the spectral weight increasing, and the green arrow shows the spectral weight decreasing. \textbf{c} The real and imaginary part of self-energy of Fe1 and Fe2 atoms in no-magnetic order. \textbf{d} The several highest probabilities of the Fe atomic states from the impurity solver in non-magnetic calculations. The $\uparrow$ arrow means $S_{z}$ is positive and the $\downarrow$ arrow means $S_{z}$ is negative. \textbf{e} The real part of self-energy of Fe1 and Fe2 atoms in ferromagnetic order. \textbf{f} The several highest probabilities of the Fe atomic states from the impurity solver in ferromagnetic order.}
\label{fig2}
\end{center}
\end{figure*}

\vspace{10pt}
\noindent \textbf{Mechanism of magnetic phase transition using DFT+DMFT calculations} \\
To further explore the electronic structures and magnetism with dynamical correlation effects, we systematically perform DFT+DMFT calculations on FGT \cite{Kotliar2006,Haule2010}. This method has been successfully applied to correlated materials such as Hund metals and heavy fermion materials \cite{Shim2007,Yin2011}. For the magnetic calculations using DFT+DMFT, we consider ten Fe-$d$ orbitals and introduce symmetry breaking of the self-energy at the initial stage. As expected, our DFT+DMFT simulations show that the system converges to non-magnetic order at high temperature in Fig.~\ref{fig2}a. The spectral function exhibits large blurred regions, and the DOS shows very broad peaks, mainly from the incoherent states of Fe atoms around the Fermi level (from -4 eV to 4 eV). This incoherence arises from the Hubbard bands of Fe atoms, originating from the strong Coulomb interaction. Moreover, the system converges to the ferromagnetic order at low temperature in Fig.~\ref{fig2}b. Compared with high-temperature results, the bands above the Fermi level become sharp in the spin-up channel, while the bands remain blurred below the Fermi level. Conversely, in the spin-down channel, the upper Hubbard bands are retained, and the bands below the Fermi level become sharp. Additionally, the bands close to the Fermi level become more distinct in both spin channels, thus forming quasiparticles, which result in multiple peaks in DOS around the Fermi level. The coherent quasiparticle states in the low-energy range and incoherent atomic-like behavior in the high-energy range have also been observed in previous theoretical studies of ferromagnetic metals, such as Fe, Co, and Ni \cite{Grechnev2007,Katsnelson1999}.

Here, we conclude several significant findings from further inspecting the microscopic processes in DFT+DMFT calculations. The most important observation is the spectral weight transfer in Hubbard bands, rather than the spin splitting suggested by the Stoner model. The colored arrows in Fig.~\ref{fig2}b indicate the primary areas where spectral weight transfer occurs. This behavior is significantly different from those observed in DFT and DFT+U calculations. Previous ARPES experiments have strongly indicated this spectral weight transfer behavior \cite{Zhuang2016}. They found that the spectral weight transfer is independent from the details of band structures, with the enhanced DOS at 500 meV and suppressed DOS at 200 meV below the Fermi energy in ferromagnetic state. Furthermore, recent ARPES measurements also show a weak temperature-dependent evolution of band shift towards the Curie temperature, suggesting that local moments may play a crucial role in ferromagnetic order \cite{Xu2020,Wu2024}. In short, our proposed mechanism of spectral weight transfer agrees well with ARPES measurements. Additionally, previous DFT+DMFT calculations of Fe and Ni also reveal a weak temperature dependence of the $d$-band shift within the temperature range from 0.6 to 0.9 $T_{c}$ \cite{Lichtenstein2001}.

\begin{table}
\caption{\label{tab1} \textbf{The mass enhancements of Fe-orbitals at paramagnetic and ferromagnetic order in FGT.} $m_{_{\rm DFT}}$ represents the mass from DFT calculations. The respective Fe-3$d$ states in FGT split into three classes: a $d_{z^2}$-like ${\rm a_{1g}}$ orbital, two degenerate ${\rm e_{g}^{,}}$ orbitals and two degenerate ${\rm e_{g}}$ orbitals. The $\uparrow$ arrow means spin-up channel and the $\downarrow$ arrow means spin-down channel.\\}
\centering
\begin{spacing}{1.05}
\begin{tabular}{c|c|c|c|c|c}
\hline
\hline
\multicolumn{3}{c|}{} & $m^{*}_{\rm a_{1g}}/m_{_{\rm DFT}}$ & $m^{*}_{\rm e_{g}^{,}}/m_{_{\rm DFT}}$ & $m^{*}_{\rm e_{g}}/m_{_{\rm DFT}}$  \\
\hline
\multirow{2}{*}{ PM } & \multicolumn{2}{c|}{Fe1} & 9.1 & 4.5 & 7.6  \\
\cline{2-6}
                               &  \multicolumn{2}{c|}{Fe2} & 1.5 & 1.8 & 2.3  \\
\hline
\multirow{4}{*}{ FM } & \multirow{2}{*}{ Fe1 } & $\uparrow$ & 1.4 & 1.4 & 1.3  \\
\cline{3-6}
                               &                                 &  $\downarrow$ & 1.3 & 1.3 & 1.2  \\
\cline{2-6}
                              & \multirow{2}{*}{ Fe2 } & \ \  $\uparrow$ \ \ & 1.4 & 1.4 & 1.3  \\
\cline{3-6}
                              &                                 & \ \  $\downarrow$ \ \  & 1.6 & 1.4 & 2.0  \\
\hline
\hline
\end{tabular}
\end{spacing}
\end{table}

Secondly, a crossover is observed from high-temperature incoherence states to low-temperature coherence states in FGT, similar to what occurs in Hund metals \cite{Yin2011,Georges2024}. It is supported by recent optical experiments, indicating that Hund's coupling is essential in FGT \cite{Corasaniti2020}. In our DFT+DMFT calculations, the formation of ferromagnetic order is more sensitive to Hund's coupling $J$ than to Coulomb repulsion $U$, highlighting the critical role of Hund's coupling in establishing ferromagnetic order (See Supplementary Note 2). The incoherent Hubbard bands originate from the significant spin scattering of Fe-$d$ electrons, which can be clearly seen from the self-energy. The imaginary parts of the self-energy in Fig.~\ref{fig2}c have significant values at zero frequency, approximately 2 eV for Fe1 and 0.1 eV for Fe2. However, at low temperatures, the imaginary parts of the self-energy approach zero, leading to the quasiparticle bands becoming distinct near the Fermi level, as shown in Fig.~\ref{fig2}b. It is important to note that the sharpness observed above the Fermi level for majority-spin states and below the Fermi level for minority-spin states is predominantly influenced by band occupation, as suggested by prior studies on excitations in cobalt \cite{Monastra2002}. The higher occupation in the spin-up channel results in small upper Hubbard bands, which causes the spectral function to appear sharp above the Fermi level. Conversely, this situation is reversed in the spin-down channel. Furthermore, correlation effects result in the renormalization of the Fe-$d$ bands near the Fermi level, which can be estimated from the real parts of self-energy with quasiparticle weight, $Z=(1-\partial_{\omega} \rm{Re}\Sigma|_{\omega=0})^{-1}$. The real parts of the self-energy are illustrated in Fig.~\ref{fig2}e, revealing characteristics indicative of moderate correlation. Table~\ref{tab1} lists the mass enhancements ($m^{*}/m_{\rm LDA}$) of each orbital, calculated as $Z^{-1}$ in DMFT. Our results demonstrate that all the Fe-$d$ orbitals are renormalized in both non-magnetic and ferromagnetic phases.

Thirdly, it is vital to shed light on the differences between FGT and other paramagnetic Hund metals \cite{Haule2008,Mravlje2011}. We focus on the detailed evolution of electronic structures during the magnetic transition. The highest atomic probabilities from the DMFT impurity solver is shown in Fig.~\ref{fig2}d. Despite the fact that the average total spin of Fe1 and Fe2 are zero in paramagnetic order, the most significant cases are $S_{z}=\pm5/2$, and they have the same probabilities. This implies that the Fe atoms have large local moments and fluctuate significantly over time. In the ferromagnetic state, the highest probabilities of Fe1 and Fe2 are primarily composed of only positive values of $S_{z}$ in Fig.~\ref{fig2}f, indicating the formation of magnetic order. Compared with paramagnetic Hund metals (See Supplementary Note 3), we recognize that the long-range magnetic order will rapidly suppress the spin fluctuations, making it easier to form quasiparticles in the magnetic phase.

Finally, we recognize that Fe1 and Fe2 atoms exhibit different physical behavior, with Fe1 being described by Mott physics and Fe2 characterized by Hund physics. This difference is evident in the histogram associated with each atomic configuration (See Supplementary Note 4). In Figs.~\ref{fig2}d and~\ref{fig2}f, Fe1 appear concentrated in a single charge state, $N=5$, typical of the Mott physics \cite{Georges2024}. However, the histogram of Fe2 extend over many charge states, a direct signature of Hund physics \cite{Yin2011}. Another piece of evidence is the feature of DOS. The partial DOS of Fe1 has a pseudogap at the Fermi level between two broad humps near $\pm$1 eV in Fig.~\ref{fig2}a, which is very similar to the Mott system V$_{2}$O$_{3}$ \cite{Deng2019}. Conversely, the partial DOS of Fe2 is characterized by a single broad feature, similar to the Hund metal Sr$_{2}$RuO$_{4}$ \cite{Deng2019}. We realize that the different physics of Fe1 and Fe2 may be due to their local environments, with the occupation of Fe1 and Fe2 being 5.0 and 5.8 in our DFT+DMFT calculations, corresponding to Fe1$^{3+}$ and Fe2$^{2+}$. In multi-orbital correlated systems, Hund's coupling has a Janus-faced effect depending on the electron occupation \cite{Medici2011}. As Fe1 is close to half-filling, an increase in Hund coupling results in a widening of the Mott gap, which is conducive to the emergence of Mott physics. In contrast, for Fe2, which is away from half-filling, the Mott gap decreases with increasing Hund coupling $J$, allowing Hund physics to play a predominant role. The valences of Fe atoms in our calculations are obtained using an exact double-counting scheme in DFT+DMFT calculations \cite{Haule2015} and are consistent with previous predictions (See Supplementary Note 5) \cite{Deng2018,Deiseroth2006}.

\begin{figure*}
\begin{center}
\includegraphics[width=0.96\textwidth]{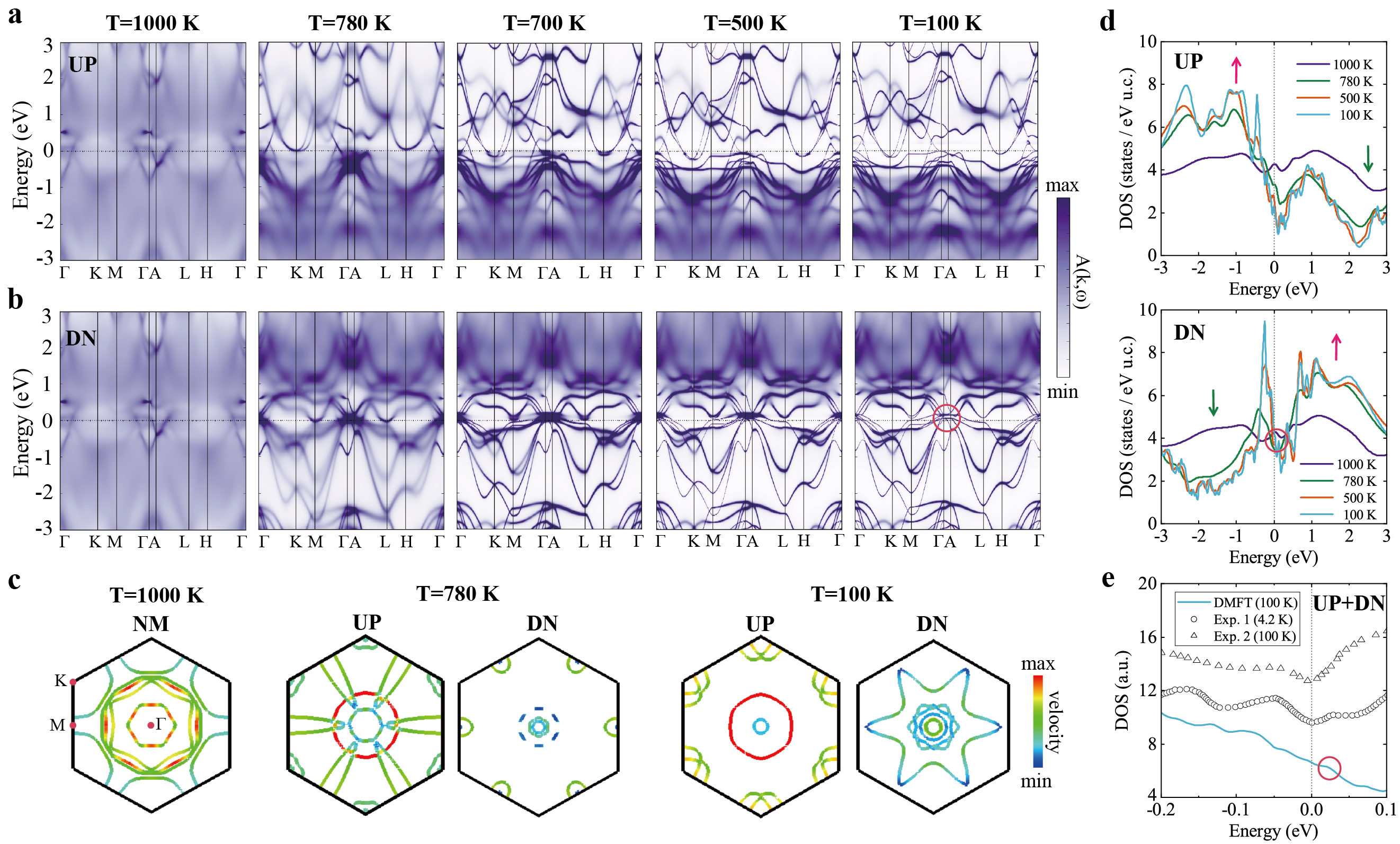}
\caption{\textbf{The results of DFT+DMFT calculations of nonstoichiometric Fe$_{2.82}$GeTe$_{2}$ with temperatures.} \textbf{a,b} The spectral functions of the spin-up and down channels with temperature from 1000 K to 100 K. \textbf{c} The DFT+DMFT Fermi surfaces at different temperatures. \textbf{d} The total density of states of spin-up and spin-down channels with varying temperatures. The colored arrows indicate the regions of spectral weight transfer. \textbf{e} The calculated DOS is compared with experimental measurements \cite{Zhang2018,Zhao2021}.}
\label{fig3}
\end{center}
\end{figure*}

\vspace{10pt}
\noindent \textbf{Effect of Fe vacancy on electronic structures} \\
Here, we summarize the common features observed in previous ARPES measurements. Notably, there are several hole-like bands across the Fermi level at the $\Gamma$ point, consisting of at least one circular-shaped pocket and one hexagonal-shaped hole pocket. Additionally, a small elliptical-shaped electron-like pocket is present at the $K$ point. Furthermore, flat bands are observed at approximately 0.5 eV below the Fermi level along the $\Gamma$-$K$ path. However, our calculations reveal a systematic discrepancy between the electronic structures of FGT and the ARPES measurements \cite{Zhang2018,Xu2020,Wu2024}. Specifically, as shown in Fig.~\ref{fig2}b, we find that the hole-like bands at the $\Gamma$ point are situated 0.5 eV below the Fermi level, which is lower than the experimental results. Moreover, the flat bands along the $\Gamma$-$K$ path are found to be approximately -1 eV in binding energy, also lower than those reported in the experiments. Although surface effects could possible partially explain the discrepancy \cite{Wang2021}, the nonstoichiometry in FGT is significant due to the nonnegligible vacancy of Fe2 atoms \cite{Deiseroth2006,Verchenko2015,Jang2020}. 

Figure~\ref{fig3} shows the temperature-dependent electronic structures in nonstoichiometric Fe$_{3-\delta}$GeTe$_{2}$ ($\delta\sim0.18$). Given that Fe deficiency results in hole doping \cite{Jang2020}, we implement hole doping simulations by reducing the total number of electrons to decrease computational costs. As expected, the main findings from stoichiometric FGT calculations are retained, including the spectral weight transfer in Hubbard bands, low-temperature quasiparticles with ferromagnetic order, and the coexistence of Hund physics and Mott physics. We observe that the band location remains relatively constant with temperature in Figs.~\ref{fig3}a and~\ref{fig3}b, which agrees well with recent APRES measurements \cite{Xu2020,Wu2024}. Furthermore, the presence of Fe vacancy induces a significant upward shift in the band structures near the Fermi level compared to the stoichiometric results (See Supplementary Note 6). Specifically, as illustrated in Fig.~\ref{fig3}c, several circular and hexagonal-shaped Fermi pockets form at the $\Gamma$ point in the nonstoichiometric calculations. Additionally, the elliptical-shaped pockets at the $K$ point closely resemble those observed in experimental results. Moreover, a complex-shaped Fermi pocket appears near the $M$ point in the spin-down channel, which is consistent with experimental findings \cite{Zhang2018,Xu2020,Wu2024}. At low temperatures, there are quasiparticle flat bands emerging in the spin-down channel near -0.5 eV, which are also observed in ARPES experiments at the same binding energy. These correspondences with experimental data underscore the significant improvements over the stoichiometric results \cite{Zhang2018,Xu2020}. Figure~\ref{fig3}d shows that spectral weight transfer occurs as temperatures decrease. Meanwhile, the DOS has many small sharp peaks due to the complex flat quasiparticle bands formed at low temperatures. The agreement between our hole-doping calculations and ARPES measurements suggests the vital role of Fe2 vacancies in FGT.

\vspace{10pt}
\noindent \textbf{Origin of heavy fermion behavior in ferromagnets} \\
Another critical issue is the emergence of heavy fermion behavior within ferromagnetic order \cite{Zhang2018}, as the two often contradict each other. Measurements obtained from scanning tunneling microscopy (STM) reveal the formation of two peaks at -35 meV and -170 meV below the Fermi level, coinciding with the transition to ferromagnetic order. Additionally, a shoulder-like peak, identified as the Kondo peak, is observed at approximately 20 meV above the Fermi level at lower temperatures \cite{Zhang2018,Zhao2021}. In our DFT+DMFT calculations for Fe$_{2.82}$GeTe$_{2}$, the presence of low-temperature ferromagnetic order significantly reduces spin fluctuations, resulting in sharp quasiparticle flat bands and corresponding prominent peaks in the DOS, as illustrated in Fig.~\ref{fig3}d. Notably, a distinct peak appears at around 20 meV above the Fermi level, indicated by the red circle in Figs.~\ref{fig3}d and~\ref{fig3}e. The emergence of these distinct peaks in the DOS near the Fermi level under ferromagnetic order is consistent with experimental observations.

Although the peak observed above the Fermi level aligns with the position found in experiments, its height is not sufficiently pronounced. Our calculations indicate that these peaks originate from the hole-like flat quasiparticle bands located at the $\Gamma$ point. Furthermore, as temperature decreases, the quasiparticle bands at the $\Gamma$ point show strong hybridization with other relatively itinerant bands, where the partial DOS of Fe1, Fe2, and Te atoms have similar shapes and peaks positions over a wide range (See Supplementary Note 7). Moreover, at low temperatures, the imaginary parts of the real-frequency hybridization functions exhibit a substantial increase near the Fermi level, particularly for the spin-down channel of the Fe1-$\rm e_{g}'$ orbital (See Supplementary Note 8). Besides, these $\Gamma$-centered hole-like bands have a small Fermi velocity in Fig.~\ref{fig3}c. Based on this evidence, we conclude that the possible hybridization between these well-defined flat bands and other itinerant bands may be the mechanism responsible for the heavy fermion behavior observed at low temperatures, similar to that seen in $f$-electron heavy fermion materials. Additionally, the crossover temperature for heavy fermion behavior is reported to be between 110 and 150 K based on experimental findings \cite{Zhang2018}. The large specific heat coefficient has been measured at lower temperatures \cite{Chen2013}. And the Kondo peak can be distinctly identified in STM experiments at 6 K \cite{Zhang2018}. However, in our DFT+DMFT calculations, we have to consider ferromagnetism and the presence of two different Fe atoms. The computations conducted at lower temperatures will exceed our computational capabilities. 

In contrast to $f$-electron heavy fermion systems, various mechanisms have been investigated in $d$-electron systems, such as geometrical frustration in LiV$_{2}$O$_{4}$ \cite{Shimizu2012}, symmetry-enforcement in CaCu$_{3}$Ir$_{4}$O$_{12}$ \cite{Liu2017}. In our study of FGT, Hund’s coupling promotes the formation of ferromagnetic order, which in turn reduces spin fluctuations and enhances the flat bands near the Fermi level. At lower temperatures, these well-defined flat bands may subsequently hybridize with other itinerant bands, leading to heavy fermion behavior. This proposed mechanism may offer a possible explanation of how heavy fermion behavior can be promoted by ferromagnetic order.

\begin{figure}[t]
\begin{center}
\includegraphics[width=0.48\textwidth]{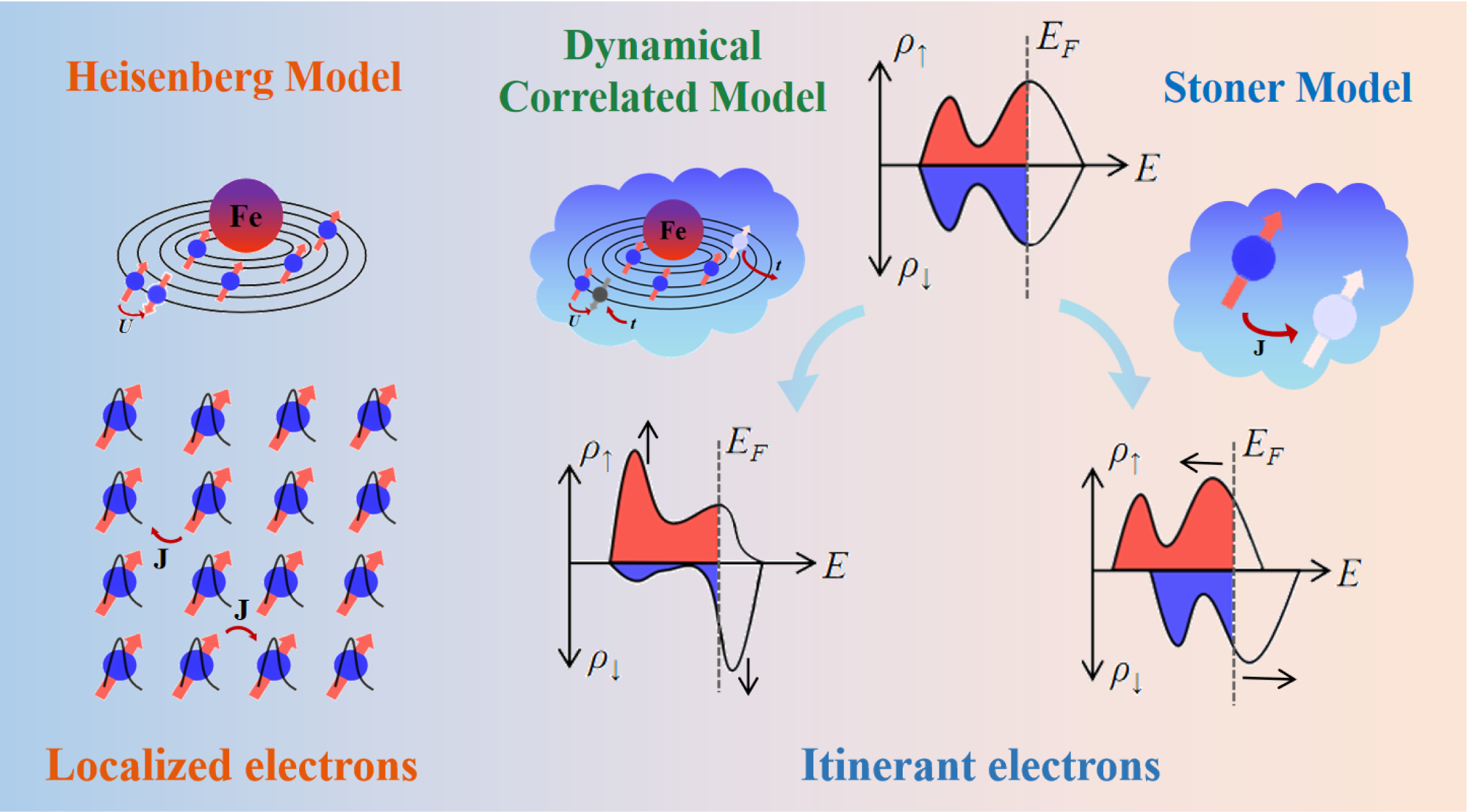}
\caption{\textbf{The schematic diagram of electronic structure evolution in itinerant ferromagnets.} The left and right sides show the local Heisenberg model and the itinerant Stoner model, respectively. In the Heisenberg model, the electrons are localized and cannot move through the crystal. In the Stoner model, electrons can move in the crystal, and the spin splitting comes from the different spin channels that moves in the opposite direction. In correlated itinerant magnets, dynamical correlation effects will lead to collective excitation, making spectral weight transfer and quasiparticle forming through magnetic phase transition.}
\label{fig4}
\end{center}
\end{figure}

\vspace{10pt}
\noindent \textbf{Discussion} \\
In strongly correlated physics, various exotic phenomena are closely related to the local moments of correlated atoms. One such outstanding material is the strongly correlated ferromagnetic metal FGT. However, the origin of magnetism in itinerant ferromagnets has been a controversial topic for several decades. As shown in Fig.~\ref{fig4}, the Heisenberg model is used to describe the origin of magnetism in insulators \cite{Heisenberg1928}. In contrast, the Stoner model, based on the mean-field approximation, provides a phenomenological description of spin splitting in magnetic metals \cite{Stoner1947}. However, the Stoner model has been proven inadequate for these correlated systems \cite{Xu2020,Wu2024}. Our magnetic DFT+DMFT calculations aim to resolve this problem from the many-body dynamical correlation point of view. In ferromagnetic metals, the dual nature of $d$-electrons, arising from correlation effects, plays a crucial role. As Hund's coupling can promote long-range ferromagnetic order, the opposing tendency in spectral weight transfer between the Hubbard bands in the two spin channels fully demonstrates the localized nature of $d$-electrons. Concurrently, the establishment of long-range ferromagnetic order enhances the formation of quasiparticle flat bands near the Fermi level through a positive feedback mechanism, which rapidly reduces spin fluctuations, thereby reflecting the itinerant characteristics of $d$-electrons. Ultimately, this leads to the simultaneous occurrence of spectral weight transfer, long-range magnetic order, and even other competing orders, such as heavy fermion behavior. 

As FGT is a layered two-dimensional vdW material, non-local correlation effects may also play a significant role. Particularly in the vicinity of the magnetic transition, long-wavelength spin waves should not be overlooked \cite{Lichtenstein2001}. A prior study stimulating the Curie temperature in Fe and Ni demonstrates that the single-site nature of the DMFT approach fails to account for the reduction of $T_{c}$ \cite{Lichtenstein2001}. This overestimation of $T_{c}$ is also evident in our calculations. Moreover, the inadequate mass renormalization observed in our results may also highlight the necessity of incorporating non-local fluctuations, as suggested by previous investigations of iron \cite{Barriga2009}. Therefore, the non-local correlation effect in FGT warrant further investigation.

In summary, through systematic magnetic DFT+DMFT calculations on FGT, we have uncovered a systematic microscopic description of electronic structures in correlated ferromagnetic metals based on dynamical correlation effects. The Hund's coupling results in long-range ferromagnetic order accompanied by spectral weight transfer between the two spin channels. This long-range magnetic order suppress spin fluctuation rapidly, leading to the formation of renormalized quasiparticle bands. Our work also show that heavy fermion behavior can be promoted by long-range magnetic order in itinerant magnets, through flat quasiparticle bands around the Fermi level further hybridizing with other itinerant bands at lower temperatures. In addition, the two Fe sites exhibit different physics, with Fe1 characterized by Mott physics and Fe2 by Hund physics, induced by the diversity of their atomic environment. Overall, our research indicates that multi-site itinerant magnets present a new frontier for discovering new quantum states and their competitions.

\vspace{10pt}
\noindent \textbf{Methods} \\
\textbf{DFT calculations}\\
The first-principles DFT calculations were performed using the projector-augmented wave (PAW) method \cite{Blochl1994}, as implemented in the Vienna ab initio simulation package (VASP) \cite{Kresse1996} with the Perdew–Burke–Ernzerhof (PBE) functional \cite{Perdew1996}. The strongly correlated correction was considered with the GGA+U method, taking the suggested correlated values as $U_{\rm eff}=$ 4 eV from previous work \cite{Zhu2016}. We also used doubly screened Coulomb correction (DSCC) approach to check the correlated values and got similar values \cite{Liu2023}. The cut-off energy of the plane wave basis-set was set to 500 eV. The Monkhorst-Pack gird was chosen as 15$\times$15$\times$3 to ensure convergence. The total energies were converged to 10$^{-8}$ eV. In our calculations, we adopted experimental lattice parameters with $a=b=3.99\,$\AA{} and $c=16.33\,$\AA{} \cite{Deiseroth2006}. 

\vspace{10pt}
\noindent \textbf{DFT+DMFT calculations}\\
The DFT+DMFT calculations were employed to treat the electronic correlations of Fe-$d$ orbitals. Full-potential linearized augmented plane-wave method implemented in the WIEN2k package was used in DFT+DMFT \cite{Blaha2014}. We employed the hybridization expansion continuous-time quantum Monte Carlo method as the impurity solver \cite{Werner2006,Haule2007} and applied the exact double-counting scheme \cite{Haule2015}. The correlation parameters were set $U=$ 5 eV and $J=$ 0.8 eV, as suggested from earlier work \cite{Zhu2016}. We used the exact double-counting scheme, and it can get the proper occupied number of Fe atoms, consistent with experiments \cite{Deng2018,Deiseroth2006}. We have examined other double-counting schemes, including the nominal and fully localized limit (FLL) schemes. Although the occupation numbers of the Fe atoms vary with different double-counting schemes, the primary finding of spectral weight transfer in our study remains consistent. The occupation numbers of different schemes are presented in Supplementary Note 5. To account for the effects of Fe vacancy and reduce computational costs, we employed hole doping simulations by decreasing the total number of electrons in accordance with the concentration of Fe vacancies, as Fe deficiency induces hole doping in this material. Previous research has confirmed that changing electron numbers leads to similar results in constructing supercells with Fe2 vacancies \cite{Jang2020,Lee2023}.  In order to perform magnetic DFT+DMFT calculations, we consider 10 Fe-$d$ orbitals and break the symmetry with a change in the real-part energy of the initial self-energy of both two types of Fe atoms.

\vspace{10pt}
\noindent        
\textbf{Acknowledgements} \\
The authors thank Yingying Cao, Beilei Liu and Kun Zhai for the fruitful discussions. This work is supported by the National Natural Science Foundation of China (Grant Nos. U2230401, 12204033, 52371174), the National Key Research and Development Program of China (Grant No. 2021YFB3501503), the Foundation of LCP, the Fundamental Research Funds for the Central Universities (Grant No. FRF-TP-22-097A1), the State Key Lab of Advanced Metals and Materials (Grant No. 2022Z-13) and the Young Elite Scientist Sponsorship Program by BAST (Grant No. BYESS2023301). We thank the Tianhe platforms at the National Supercomputer Center in Tianjin. Numerical computations were also performed on Hefei advanced computing center.

\vspace{10pt}
\noindent        
\textbf{Author contributions} \\
F.Y.T. and H.F.S. conceived the idea and supervised the project; Y.J.X. and Y.C.W. performed the calculations; X.T.J., Y.L. and H.F.L. participated in the discussions. Y.J.X.,F.Y.T. and H.F.S. wrote the paper with input from all authors. All authors contributed to the interpretation of the data and revised the paper. Y.J.X. and Y.C.W. contributed equally to this work.

\vspace{10pt}
\noindent        
\textbf{Competing interests} \\
The authors declare no competing interests.

\vspace{10pt}
\noindent        
\textbf{Data availability} \\
Data are available upon request.

\end{document}